\documentclass[aps,prc,twocolumn,showpacs,superscriptaddress,floatfix,10pt]{revtex4-1}
\usepackage{graphicx}
\usepackage{dcolumn}
\usepackage{bm}
\usepackage{xcolor}
\usepackage{amsfonts}
\usepackage{amssymb}
\usepackage{amsmath}
\usepackage[colorlinks,linkcolor=blue]{hyperref}

\begin{document}
\title{Two-body dissipation effects on synthesis of superheavy elements}

\author{M. Tohyama}
\affiliation{Kyorin University School of Medicine, Mitaka, Tokyo 181-8611, Japan}
\author{A. S. Umar}
\affiliation{Department of Physics and Astronomy, Vanderbilt University, Nashville, Tennessee 37235, USA} 
 

\begin{abstract}
To investigate the two-body dissipation effects on the synthesis of superheavy elements, we calculate
low-energy collisions of the $N=50$ isotones ($^{82}$Ge, $^{84}$Se, $^{86}$Kr and $^{88}$Sr) on $^{208}$Pb
using the time-dependent density-matrix theory (TDDM). TDDM is an extension of the time-dependent
Hartree-Fock (TDHF) theory and can determine the time evolution of 
one-body and two-body density matrices. Thus TDDM describes both one-body and two-body dissipation of collective energies. 
It is shown that the two-body dissipation may increase fusion cross sections and enhance the synthesis of superheavy elements.
\end{abstract}
\maketitle

\section{introduction}
The creation of new elements is one of the most novel and challenging research
areas of nuclear physics~\cite{armbruster1985,hofmann1998,hofmann2000,oganessian2007}.
The search for a region of the nuclear chart that can sustain the so
called \textit{superheavy elements} (SHE) has led to intense experimental activity
resulting in the discovery and confirmation of elements with atomic numbers as large as
$Z=118$~\cite{oganessian2010,oganessian2012,khuyagbaatar2014}.
The theoretically predicted \textit{island of stability}
in the SHE region of the nuclear chart is the result of new
proton and neutron shell-closures, whose location is not precisely
known~~\cite{bender1999,staszczak2013,cwiok2005}.
The experiments to discover these new elements are notoriously difficult, with
fusion evaporation residue (ER) cross-section in pico-barns.

The time-dependent Hartree-Fock theory (TDHF) provides us with a microscopic and self-consistent way to study nuclear 
dynamics and has extensively been used to study low-energy heavy-ion collisions~\cite{negele1982,simenel2012}.
However, in such calculations approximations of any type limit the number of degrees of freedom accessible during a collision,
and hence the nature and degree of dissipation~\cite{umar1986a,reinhard1988,umar1989,umar2006c}. The understanding of
the dissipative mechanisms in the TDHF theory is vital for establishing the region of validity of the
mean-field approximation and providing estimates for the importance of the mean-field effects at
higher energies. In TDHF, the dissipation of the translational kinetic energy of the two ions is due to
the collisions of single particle states with the walls of the time-dependent potential. This leads to the
randomization of the motion characterized by the distribution of energy among all possible degrees
of freedom of the system. The complete equilibration of the translational kinetic energy among all
possible degrees of freedom is commonly accepted as being the definition of fusion whereas the
incomplete equilibration results in inelastic collisions.

Recently TDHF simulations 
using symmetry unrestricted three-dimensional codes with full Skyrme effective interactions~\cite{umar2006c,maruhn2014}
have been reported for fusion reactions~\cite{umar2010a,guo2012,simenel2013a}, particle transfer 
reactions~\cite{simenel2010,simenel2011,sekizawa2013,sekizawa2015}, quasi-fission 
processes~\cite{simenel2012b,simenel2012c,oberacker2014,umar2015c},calculation of ion-ion interaction 
potentials~\cite{umar2006a,oberacker2010,umar2010a,umar2014a}, 
and to dynamics of fission~\cite{simenel2014a,goddard2015,scamps2015}.
Even though TDHF simulations have become sophisticated, it is still plausible that additional dissipation of collective energies due to  
two-body mechanism plays an important role in critical situations like fusion processes. 
In this paper we study possible effects of the two-body dissipation on the synthesis of superheavy elements
using the time-dependent density-matrix theory (TDDM)~\cite{wang1985,gong1990,tohyama1985,tohyama2002b}. 
TDDM which is
formulated by truncating the Bogoliubov-Born-Green-Kirkwood-Yvon (BBGKY) hierarchy 
for reduced density matrices at a two-body level consists of the coupled equations of motion for one-body and two-body density matrices. The two-body
dissipation is included through the coupling to the two-body density matrix.
We consider so-called cold fusion~\cite{hofmann2000} using $^{208}$Pb as the target and 
the $N=50$ isotones ($^{82}$Ge, $^{84}$Se, $^{86}$Kr and $^{88}$Sr) as the projectiles. 
The nuclei $^{82}$Ge and $^{84}$Se are unstable but included to study the charge dependence of fusion reactions. Such unstable projectiles 
may be realized as radioactive beams~\cite{loveland2007}.
We show that the two-body dissipation could enhance the synthesis of superheavy elements.

The paper is organized as follows. A brief outline of the TDDM formalism in connection to TDHF is given in Sec.~\ref{sec.formalism}. 
Calculational details are given in Sec.~\ref{sec.compute}.
Results are discussed in Sec.~\ref{sec.results}, followed by the conclusion in Sec.~\ref{sec.summary}.

\section{Formulation}
\label{sec.formalism}
Here we give a brief outline of the TDDM formalism. Further details can be found in~\cite{gong1990}.
We start with a many-body Hamiltonian $H$ consisting of a one-body part and a two-body interaction
\begin{eqnarray}
H=\sum_{\alpha\alpha'}\langle \alpha|t|\alpha'\rangle a^\dag_\alpha a_{\alpha'}
+\frac{1}{2}\sum_{\alpha\beta\alpha'\beta'}\langle\alpha\beta|v|\alpha'\beta'\rangle
a^\dag_{\alpha}a^\dag_\beta a_{\beta'}a_{\alpha'},
\nonumber \\
\label{totalH}
\end{eqnarray}
where $a^\dag_\alpha$ and $a_\alpha$ are the creation and annihilation operators of a particle at
a time-dependent single-particle state $\alpha$.
TDDM gives the coupled equations of motion for the one-body density matrix (the occupation matrix) $n_{\alpha\alpha'}$
and the correlated part of the two-body density matrix $C_{\alpha\beta\alpha'\beta'}$.
These matrices are defined as
\begin{eqnarray}
n_{\alpha\alpha'}(t)&=&\langle\Phi(t)|a^\dag_{\alpha'} a_\alpha|\Phi(t)\rangle,
\\
C_{\alpha\beta\alpha'\beta'}(t)&=&\langle\Phi(t)|a^\dag_{\alpha'}a^\dag_{\beta'}
 a_{\beta}a_{\alpha}|\Phi(t)\rangle
\nonumber \\
&-&(n_{\alpha\alpha'}(t)n_{\beta\beta'}(t)-n_{\alpha\beta'}(t)n_{\beta\alpha'}(t)) ,
 \label{rho2}
\end{eqnarray}
where $|\Phi(t)\rangle$ is the time-dependent total wavefunction
$|\Phi(t)\rangle=\exp[-iHt/\hbar] |\Phi(t=0)\rangle$.
The single-particle wavefunctions $\phi_\alpha$ satisfy a TDHF-like equation
\begin{eqnarray}
i\hbar\frac{\partial \phi_\alpha}{\partial t}=h\phi_\alpha,
\label{mf1}
\end{eqnarray}
where 
\begin{eqnarray}
\langle\alpha|h|\alpha'\rangle=\langle \alpha|t|\alpha'\rangle
+\sum_{\lambda_1\lambda_2}
\langle\alpha\lambda_1|v|\alpha'\lambda_2\rangle_A 
n_{\lambda_2\lambda_1}.
\label{mf2}
\end{eqnarray}
Here the subscript $A$ means that the corresponding matrix is antisymmetrized.
The equations of motion for $n_{\alpha\alpha'}$ and $C_{\alpha\beta\alpha'\beta'}$ are written as~\cite{gong1990}
\begin{eqnarray}
i\hbar \dot{n}_{\alpha\alpha'}
&=&\sum_{\lambda_1\lambda_2\lambda_3}
[\langle\alpha\lambda_1|v|\lambda_2\lambda_3\rangle C_{\lambda_2\lambda_3\alpha'\lambda_1}
\nonumber \\
&-&C_{\alpha\lambda_1\lambda_2\lambda_3}\langle\lambda_2\lambda_3|v|\alpha'\lambda_1\rangle],
\label{n}
\end{eqnarray}
\begin{eqnarray}
i\hbar\dot{C}_{\alpha\beta\alpha'\beta'}
&=&B_{\alpha\beta\alpha'\beta'}+P_{\alpha\beta\alpha'\beta'}+H_{\alpha\beta\alpha'\beta'}.
\label{N3C2}
\end{eqnarray}
The matrix $B_{\alpha\beta\alpha'\beta'}$ in Eq. (\ref{N3C2}) does not contain $C_{\alpha\beta\alpha'\beta'}$ and describes 
two particle - two hole (2p-2h) and 2h-2p
excitations:
\begin{eqnarray}
B_{\alpha\beta\alpha'\beta'}&=&\sum_{\lambda_1\lambda_2\lambda_3\lambda_4}
\langle \lambda_1\lambda_2|v|\lambda_3\lambda_4 \rangle_A
\nonumber \\ 
&\times&[(\delta_{\alpha\lambda_1}-n_{\alpha\lambda_1})(\delta_{\beta\lambda_2}-n_{\beta\lambda_2})
n_{\lambda_3\alpha'}n_{\lambda_4\beta'}
\nonumber \\
&-&n_{\alpha\lambda_1}n_{\beta\lambda_2}(\delta_{\lambda_3\alpha'}-n_{\lambda_3\alpha'})
(\delta_{\lambda_4\beta'}-n_{\lambda_4\beta'})].
\nonumber \\
\label{B^0}
\end{eqnarray}
Particle - particle and hole-hole correlations 
are described by $P_{\alpha\beta\alpha'\beta'}$:
\begin{eqnarray}
P_{\alpha\beta\alpha'\beta'}&=&\sum_{\lambda_1\lambda_2\lambda_3\lambda_4}
\langle \lambda_1\lambda_2|v|\lambda_3\lambda_4 \rangle
\nonumber \\ 
&\times&[(\delta_{\alpha\lambda_1}\delta_{\beta\lambda_2}
-\delta_{\alpha\lambda_1}n_{\beta\lambda_2}
-n_{\alpha\lambda_1}\delta_{\beta\lambda_2})
{C}_{\lambda_3\lambda_4\alpha'\beta'}
\nonumber \\
&-&(\delta_{\lambda_3\alpha'}\delta_{\lambda_4\beta'}
-\delta_{\lambda_3\alpha'}n_{\lambda_4\beta'}
-n_{\lambda_3\alpha'}\delta_{\lambda_4\beta'})
{C}_{\alpha\beta\lambda_1\lambda_2}].
\nonumber \\
\label{P^0}
\end{eqnarray}
The matrix $H_{\alpha\beta\alpha'\beta'}$ contains the  particle-hole correlations:
\begin{eqnarray}
H_{\alpha\beta\alpha'\beta'}&=&\sum_{\lambda_1\lambda_2\lambda_3\lambda_4}
\langle \lambda_1\lambda_2|v|\lambda_3\lambda_4 \rangle_A
\nonumber \\ 
&\times&[\delta_{\alpha\lambda_1}(n_{\lambda_3\alpha'}{C}_{\lambda_4\beta\lambda_2\beta'}
-n_{\lambda_3\beta'}{C}_{\lambda_4\beta\lambda_2\alpha'})
\nonumber \\
&+&\delta_{\beta\lambda_2}(n_{\lambda_4\beta'}{C}_{\lambda_3\alpha\lambda_1\alpha'}
-n_{\lambda_4\alpha'}{C}_{\lambda_3\alpha\lambda_1\beta'})
\nonumber \\
&-&\delta_{\alpha'\lambda_3}(n_{\alpha\lambda_1}{C}_{\lambda_4\beta\lambda_2\beta'}
-n_{\beta\lambda_1}{C}_{\lambda_4\alpha\lambda_2\beta'})
\nonumber \\
&-&\delta_{\beta'\lambda_4}(n_{\beta\lambda_2}{C}_{\lambda_3\alpha\lambda_1\alpha'}
-n_{\alpha\lambda_2}{C}_{\lambda_3\beta\lambda_1\alpha'})].
\nonumber \\
\label{H^0}
\end{eqnarray}
The total number of particles $A=\sum_{\alpha}n_{\alpha\alpha}$ is conserved as easily understood
by taking the trace of Eq. (\ref{n}). Formally
Eqs. (\ref{n}) and (\ref{N3C2}) also conserve the total energy $E_{\rm tot}$ given by~\cite{gong1990}
\begin{eqnarray}
E_{\rm tot}&=&\sum_{\alpha\alpha'}\langle \alpha|t|\alpha'\rangle n_{\alpha'\alpha}
\nonumber \\
&+&\frac{1}{2}\sum_{\alpha\beta\alpha'\beta'}\langle\alpha\beta|v|\alpha'\beta'\rangle
(n_{\alpha'\alpha}n_{\beta'\beta}-n_{\alpha'\beta}n_{\beta'\alpha}
\nonumber \\
&+&C_{\alpha'\beta'\alpha\beta}).
\end{eqnarray}

\section{Calculational details}
\label{sec.compute}
Since our interest here is not in quantitative analysis of production rates of super-heavy elements but in 
exploration of possible effects of the two-body dissipation on their synthesis,
we consider only the head-on collisions using the TDDM code~\cite{tohyama2002b,tohyama2002a} which was developed
based on the TDHF code~\cite{umar1986a} with axial symmetry restriction. 
The assumption of the axial symmetry is justified for the head-on collisions.
We consider the collisions of the $N=50$ isotones ($^{82}$Ge, $^{84}$Se, $^{86}$Kr and $^{88}$Ge) on $^{208}$Pb
so that the total system has the charge $114 \le Z \le 120$. Although Nuclei $^{82}$Ge and $^{84}$Se are unstable, they are included in the calculations
to cover the total charges $Z=114$ and $116$. The HF ground state is used as 
the initial ground states of the colliding nuclei. In the case of the projectiles which are open-shell nuclei
it is assumed in the HF iteration process that the lowest-energy proton single-particle states in the $Z=28-40$ subshell 
are fully occupied by the corresponding number of valence protons.
The projectile nuclei thus prepared have slight deformation because 
not all single-particle states with different magnetic quantum numbers are equally occupied.
The mesh sizes used in the TDHF code are $\Delta r=\Delta z=0.5$ fm and the mesh points are $N_r\times N_z=30\times 90$. The time step size
is $\Delta t=0.75$ fm/c. We use the Skyrme III force~\cite{beiner1975} for the mean-field Hamiltonian Eq. (\ref{mf1}). Since the Skyrme III has 
a large effective mass ($m^*/m\approx 0.9$), it is possible to obtain several bound single-particle states above the Fermi level 
which are needed to define $C_{\alpha\beta\alpha'\beta'}$. 
Since the number of $C_{\alpha\beta\alpha'\beta'}$ increases rapidly with increasing number of the single-particle states, we
are forced to use a quite limited number of the single-particles states for the calculation of $C_{\alpha\beta\alpha'\beta'}$.
To solve Eqs. (\ref{n}) and (\ref{N3C2}),
we take about 20 bound single states near the Fermi level both for protons and neutrons: The number depends on the projectile nucleus.
As the residual interaction in Eqs. (\ref{n}) and (\ref{N3C2}), which should in principle be consistent with the effective interaction used for the
mean-field potential, we use a simple contact interaction  
$v({\bm r}-{\bf r'})=v_0\delta^3({\bm r}-{\bf r'})$ with $v_0=-500$ MeV~fm$^3$ to facilitate the time-consuming calculations of
the matrix elements $\langle\alpha\beta|v|\alpha'\beta'\rangle$ at each time step. 
The value $v_0=-500$ MeV~fm$^3$ is similar to the strength of the contact interactions used in the study of the pairing correlations in
tin isotopes~\cite{sandulescu2008}.
We consider that the system fuses when the colliding nuclei stick together beyond $T_{\rm f}=4000$ fm/c.
This criterion for fusion seems reasonable as compared with the TDHF fusion study by Guo and Nakatsukasa~\cite{guo2012}
for a similar heavy system $^{70}$Zn$+$$^{208}$Pb.

\section{Results}
\label{sec.results}
The results for each projectile nucleus are summarized as follows:

i) $^{82}$Ge: The total charge of this system is $Z=114$. In TDHF fusion occurs in the two different energy regions, $300$ MeV $\le E_{\rm cm}\le 380$ MeV 
and $470$ MeV $\le E_{\rm cm}\le 620$ MeV, where $E_{\rm cm}$ is the incident energy in the center-of-mass frame.
Since the system barely escapes fusion in TDHF below and above the energy ranges, 
fusion occurs in TDDM in wider energy region $285$ MeV $\le E_{\rm cm}\le 650$ MeV.
The Coulomb barrier for the system $^{76}$Ge$+$$^{208}$Pb which was estimated by Smola$\acute{\rm n}$czuk 
using the folding potential~\cite{smola2008}
is $257.5$ MeV. If a similar value of the Coulomb barrier is applied to the system $^{82}$Ge$+$$^{208}$Pb, the lowest energy for fusion in TDDM 
is about 28 MeV
larger than the Coulomb barrier, which corresponds to so-called extra push. 
The above result shows that the extra push becomes smaller due to the two-body dissipation.
\begin{figure}[!htb]
\includegraphics*[width=8.6cm]{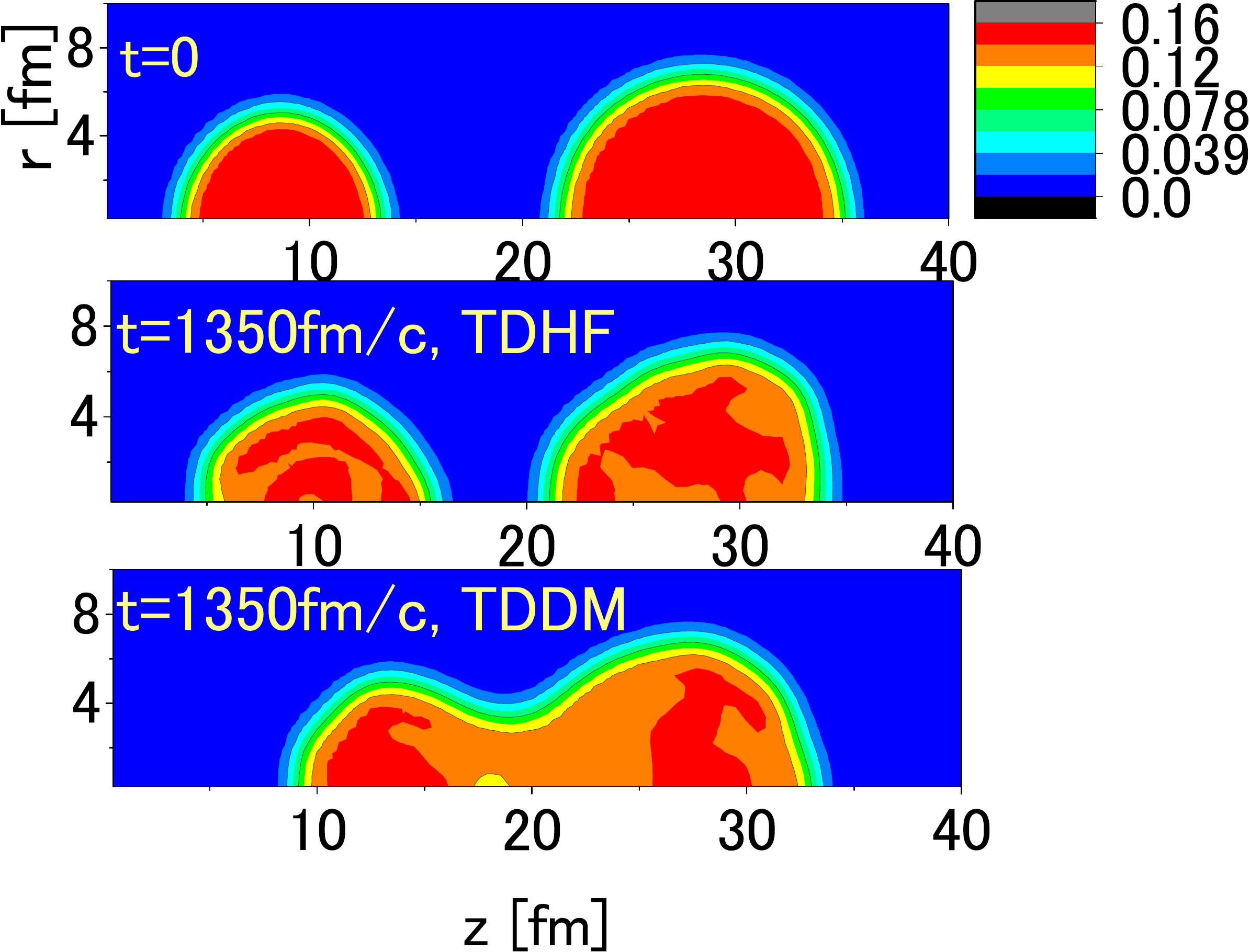}
\caption{\protect Contour plot of the density $\rho(z,r)$ [fm$^{-3}$] in the head-on collision of $^{82}$Ge$+$$^{208}$Pb at $E_{\rm cm}=450$ MeV. 
The upper part shows the density profile at $t=0$, the middle part that in TDHF at $t=1350$ fm/c and the lower part that in TDDM at $t=1350$ fm/c.
The system fuses in TDDM at this incident energy.}
\label{fusion1} 
\end{figure}
The density profiles in TDHF and TDDM at $E_{\rm cm}=450$ MeV are shown in Fig. \ref{fusion1}.
In TDDM the system fuses whereas a projectile-like fragment appears on the left-hand side in TDHF after a rather large contact period. This process in TDHF 
may
correspond to quasi fission.
The collision pattern in TDHF changes at higher incident energies around $E_{\rm cm}=600$ MeV beyond which a projectile-like 
profile appears on the right-hand side in the final state.
Therefore, the fused system in TDHF and TDDM above $E_{\rm cm}=600$ MeV has a shape similar to 
the lower part of Fig. \ref{fusion1} but reflected with respect to the plane perpendicular to the $z$ axis. 
\begin{figure}[!htb]
\includegraphics*[width=8.6cm]{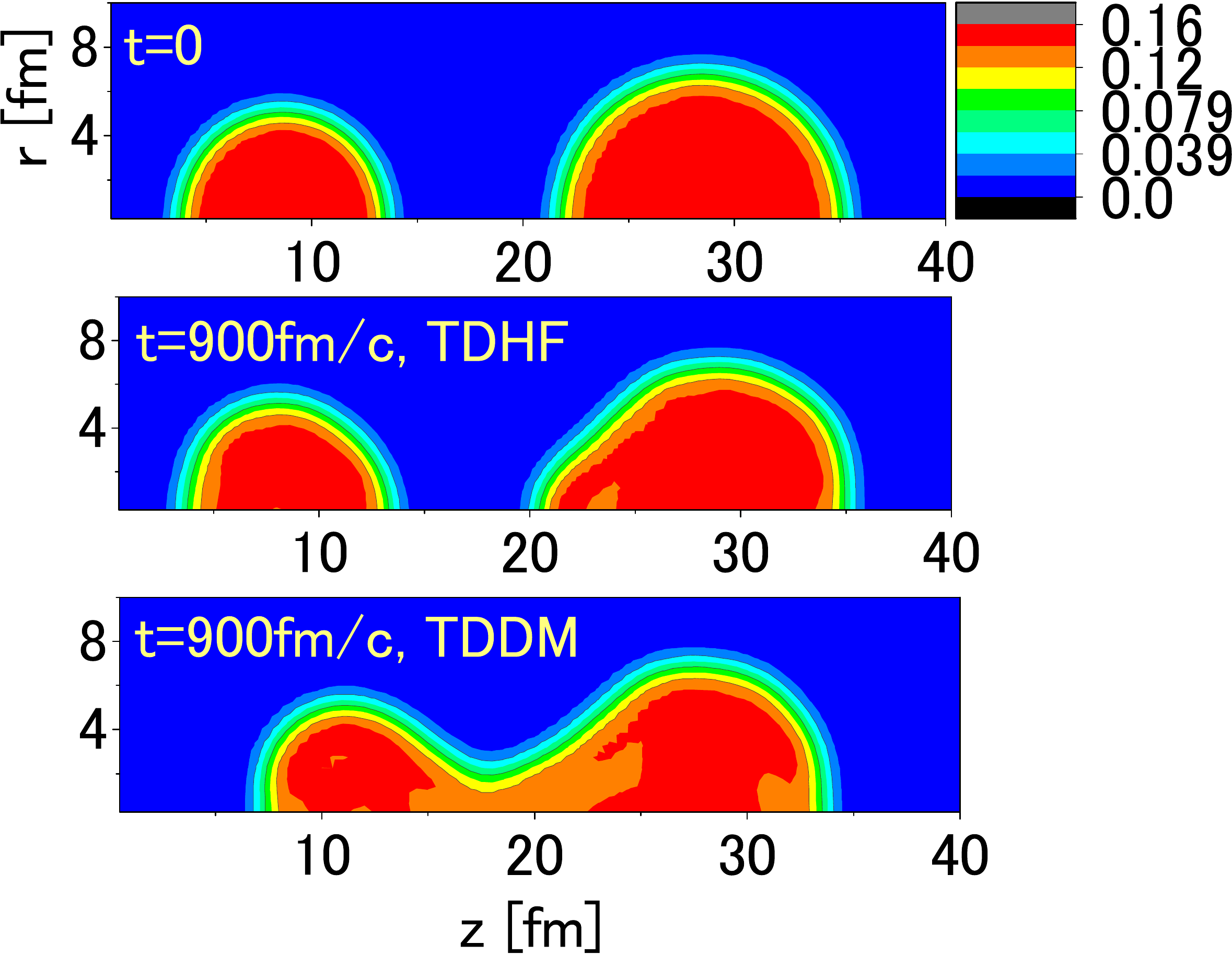}
\caption{\protect Contour plot of the density $\rho(z,r)$ [fm$^{-3}$] in the head-on collision of $^{88}$Sr$+$$^{208}$Pb at $E_{\rm cm}=340$ MeV. 
The upper part shows the density profile at $t=0$, the middle part that in TDHF at $t=900$ fm/c and the lower part that in TDDM at $t=900$ fm/c.}
\label{fusion2} 
\end{figure}

ii) $^{84}$Se : The total charge of this system is $Z=116$. 
The system fuses in TDHF in the quite narrow energy region $E_{\rm cm}= 299\pm 1$ MeV, while
fusion occurs in TDDM in the wider energy range $300$ MeV $\le E_{\rm cm}\le 400$ MeV.
The fusion threshold $E_{\rm cm}\approx 300$ MeV for $^{84}$Se$+$$^{208}$Pb is about 30 MeV
higher than the Coulomb barriers given by Smola$\acute{\rm n}$czuk~\cite{smola2008,smola2010}. 

iii) $^{86}$Kr: The total charge of this system is $Z=118$. 
The system does not fuse in TDHF although the contact time of the colliding nuclei becomes large with increasing 
incident energy: It is about $2300$ fm/c for $^{86}$Kr$+$$^{208}$Pb at $E_{\rm cm}= 650$ MeV. We cannot find 
a high-energy fusion region around $E_{\rm cm}= 600$ MeV which has been predicted by an early TDHF calculation~\cite{davies1980} for $^{84}$Kr+$^{209}$Bi.
This may be explained by the fact that they defined fusion using a smaller $T_{\rm f}\approx 1350$ fm/c.
The system fuses in TDDM in the narrow energy range $E_{\rm cm}= 341\pm 1$ MeV.
The fusion threshold $E_{\rm cm}\approx 340$ MeV for $^{86}$Kr$+$$^{208}$Pb is about 54 MeV
larger than the Coulomb barriers given by
Smola$\acute{\rm n}$czuk~\cite{smola2008,smola2010}.
The value of extra push is about half of the prediction of the Swiateki's macroscopic model~\cite{swiatecki1982} for the corresponding 
effective fissility 
$(Z^2/A)_{\rm eff}=4Z_1Z_2/A_1^{1/3}A_2^{1/3}(A_1^{1/3}+A_2^{1/3})$, where $Z_1$, $Z_2$, $A_1$ and $A_2$ are
proton and mass numbers of the colliding partners, but about twice larger than
the result of the TDHF calculation by Guo and Nakatsukasa~\cite{guo2012} for the system
$^{100}$Sn$+^{132}$Sn which has similar effective fissility.

iv) $^{88}$Sr: The total charge of this system is $Z=120$. Fusion does not occurs both in TDHF and TDDM.
In TDDM the two fragments are further slowed down than in TDHF due to the two-body dissipation as shown in Fig. \ref{fusion2} for 
$^{88}$Sr$+$$^{208}$Pb at $E_{\rm cm}=340$ MeV.  
The fact that the system does not fuse in TDDM may be due to the
truncation of the single-particle space to define $C_{\alpha\beta\alpha'\beta'}$. 
Since it is hard to increase the number of the single-particle states, 
we performed a TDDM calculation using a stronger residual interaction with $v_0=-1000$ MeV~fm$^3$ at $E_{\rm cm}=340$ MeV
and found that the system fuses. More elaborate calculations are needed for this system to conclude whether the system fuses or not in TDDM.

\section{Summary}
\label{sec.summary}
In summary,
low-energy head-on collisions of the $N=50$ isotones ($^{82}$Ge, $^{84}$Se, $^{86}$Kr and $^{88}$Sr) on $^{208}$Pb
were studied using the time-dependent density-matrix theory (TDDM). TDDM is an extension of the time-dependent
Hartree-Fock theory (TDHF) and can include the effects of the two-body dissipation which is missing in TDHF.
It was shown that the two-body dissipation expands the fusion energy range for $^{84}$Se$+^{208}$Pb and makes it possible for $^{86}$Kr$+^{208}$Pb
to fuse. Thus the two-body dissipation could play an important role in the synthesis of superheavy elements. 
The obtained results encourage further studies of the two-body dissipation effects based on the TDDM approach,
though various refinements such as increase of the single-particle space and improvement of the residual interaction are 
needed to obtain more quantitative results. 

\section*{Acknowledgments}
This work was supported in part by DOE grant No. DE-SC0013847 with Vanderbilt University.

\bibliography{VU_bibtex_master.bib}

\end{document}